\documentclass[%
 reprint,
superscriptaddress,
 amsmath,amssymb,
 aps,
floatfix,
]{revtex4-1}

\usepackage{graphicx}% Include figure files
\usepackage{caption}
\usepackage{subcaption}
\usepackage{dcolumn}% Align table columns on decimal point
\usepackage{bm}% bold math
\usepackage{float}

\begin{document}

\preprint{APS/123-QED}

\title{Encoding Gaussian curvature in glassy and elastomeric liquid crystal
polymer networks}%

\author{Cyrus Mostajeran}
\affiliation{%
Department of Engineering, University of Cambridge, Cambridge CB2 1PZ, United
Kingdom }%
\author{Taylor H. Ware}
\affiliation{
Materials and Manufacturing Directorate, Air Force Research Laboratory, Wright-Patterson Air Force Base, OH 45433, USA
}%
\affiliation{
Azimuth Corporation, Beavercreek, OH USA
}%
\author{Timothy J. White}
\affiliation{
Materials and Manufacturing Directorate, Air Force Research Laboratory, Wright-Patterson Air Force Base, OH 45433, USA
}%

\date{\today}% It is always \today, today,

\begin{abstract}

Considerable recent attention has been given to the study of shape
formation using modern responsive materials that can be preprogrammed to undergo
spatially inhomogeneous local deformations. In particular, nematic liquid crystal 
polymer networks offer exciting possibilities in
this context. In
this Letter, we discuss the generation of Gaussian curvature in thin nematic sheets using smooth
in-plane director fields patterned across the surface. We highlight specific
patterns which encode constant Gaussian curvature of prescribed sign and magnitude and present
experimental results which appear to support the theoretical predictions.
Specifically, we provide experimental evidence for the realization of positive
and negative Gaussian curvature in glassy and elastomeric liquid crystal polymer
networks through the stimulation of smoothly varying in-plane director fields.
\end{abstract}
 
\maketitle 
 
%\tableofcontents

It is well known that inhomogeneous local deformations such as differential
growth of thin elastic sheets can lead to the formation of Gaussian curvature
and complex shape transitions \cite{dervaux2008morphogenesis,Mostajeran2015}.
Modern responsive materials that can be preprogrammed to undergo prescribed spatially
inhomogeneous expansions and contractions in response to external stimuli offer
exciting possibilities for the design and production of switchable surfaces for
use in a variety of applications
\cite{klein2007shaping,kim2012designing,modes2011gaussian,aharoni2014geometry}.
Nematic liquid crystalline glasses and elastomers are particularly promising candidates
for the responsive material of
choice. Liquid crystalline polymer networks consist of long, semiflexible molecular 
crosslinked chains that possess mesomorphic order. Below certain critical
temperatures, the material may possess one-dimensional order where
the rod-like molecular elements are locally aligned about the
director $\boldsymbol{n}$ and the material is said to be in the nematic phase. 
Liquid crystalline solids experience local deformations in response to light,
heat, \textit{p}H, and other stimuli that change the molecular order. Of particular
interest are nematic glasses \cite{van2007glassy} and elastomers
\cite{warner2003liquid}, both of which have 
spontaneous deformation tensors of the
form
\begin{equation}
F=(\lambda-\lambda^{-\nu})\boldsymbol{n}\otimes\boldsymbol{n}+\lambda^{-\nu}\,\mathrm{Id}_3,
\end{equation}
where $\mathrm{Id}_3$ denotes the identity operator on $\mathbb{R}^3$.
This describes a local scaling by $\lambda<1$ along the director
$\boldsymbol{n}$ and a scaling by $\lambda^{-\nu}$ perpendicular to $\boldsymbol{n}$.
The parameter $\nu$ is known as the opto-thermal Poisson
ratio and relates the perpendicular and parallel responses
\cite{modes2011blueprinting}.

In the seminal work \cite{aharoni2014geometry}, Aharoni \emph{et al.} describe the interplay between the nematic director 
field of a thin elastomeric sheet and the resulting 3D configuration attained upon heating. In particular, they consider the 
reverse problem of constructing a director field that induces a specified 2D intrinsic geometry. 
In this paper, we follow the presentation in \cite{Mostajeran2015} and
consider 2D in-plane director field patterns on thin nematic sheets. It is
assumed that the director field does not vary across the thickness of the sheet
so that the same pattern is repeated at each level of thickness. For
sufficiently thin sheets, stimulation of the system will result in pure bending
of the sheet at no stretch energy cost and one expects an isometric immersion of
the prescribed local deformations as determined by the director field pattern.

Let $(x_1, \, x_2) \in \omega\subset \mathbb{R}^2$ be Cartesian coordinates
parametrising the mid-surface of the initially flat sheet and
$\boldsymbol{n}(x_1,x_2)=n_1\hat{\boldsymbol{e}}^1+n_2\hat{\boldsymbol{e}}^2$ be
the director field pattern across the surface, where
$\hat{\boldsymbol{e}}^1$, $\hat{\boldsymbol{e}}^2$ form the standard orthonormal
basis of $\mathbb{R}^2$. The associated in-plane spontaneous deformation tensor
$F$ has components
$
F_{\alpha\beta}=\left(\lambda-\lambda^{-\nu}\right)n_{\alpha}n_{\beta}+\lambda^{-\nu}\delta_{\alpha\beta},
$
where $\alpha, \, \beta = 1,\, 2$. The resulting 2D metric of the deformed sheet
upon stimulation is $a=F^TF$, which simplifies to
\begin{equation}
a_{\alpha\beta}=\left(\lambda^2-\lambda^{-2\nu}\right)n_{\alpha}
n_{\beta}+\lambda^{-2\nu}\delta_{\alpha\beta}.
\end{equation}
We characterise the 2D director field by an angle scalar field
$\psi=\psi(x_1,x_2)$ which specifies the in-plane orientation of the director at
each point on the initially flat sheet, so that $n_1=\cos\psi$ and
$n_2=\sin\psi$. By the \emph{Theorema Egregium} of Gauss, the Gaussian curvature
$K$ of a surface is an intrinsic geometric property that is determined by the
first fundamental form $a_{\alpha\beta}$ of the surface via 
\begin{align}
K=-\frac{1}{a_{11}}\Big(\partial_1\Gamma^2_{12}&-\partial_{2}\Gamma^2_{11}+\Gamma^1_{12}\Gamma^2_{11}\nonumber
\\
&-\Gamma^1_{11}\Gamma^2_{12}+\Gamma^2_{12}\Gamma^2_{12}-\Gamma^2_{11}\Gamma^2_{22}\Big),
\end{align}
where
$\Gamma^{\gamma}_{\alpha\beta}=\frac{1}{2}a^{\gamma\sigma}(\partial_{\alpha}a_{\sigma\beta}
+\partial_{\beta}a_{\alpha\sigma}-\partial_{\sigma}a_{\alpha\beta})$ in Einstein notation.

The Gaussian curvature determined by the nematic metric can be expressed in terms of the alignment angle field $\psi$ as
\begin{align}
K = \frac{1}{2}&\left(\lambda^{2\nu}-\lambda^{-2}\right)\left[
\left(\partial^2_2 \psi -\partial^2_1\psi -4 \partial_1 \psi \partial_2 \psi\right)\sin (2 \psi)\right. \nonumber \\
&+ \;\;
\left.2\left(\partial_1\partial_2\psi+(\partial_2\psi)^2-(\partial_1\psi)^2\right)\cos (2 \psi)\right].
\end{align}
We note here that if we rotate the director associated with a given pattern by
$\pi/2$ at every point, so that $\psi\rightarrow\psi+\pi/2$, then the resulting
Gaussian curvature flips sign at every point, since
$\sin2\psi\rightarrow-\sin2\psi$ and $\cos 2\psi\rightarrow -\cos 2\psi$. That
is, 
\begin{equation} {\label{dual}}
K\rightarrow -K \;\;\; \mathrm{as} \;\;\; \psi \rightarrow \psi+\pi/2.
\end{equation}
We refer to a pair of director field patterns that are related by a $\pi/2$
radian rotation of the directors as orthogonal duals.

\begin{figure*}
\captionsetup{justification=raggedright,
singlelinecheck=false
}
\centering
\includegraphics[width=0.80\linewidth]{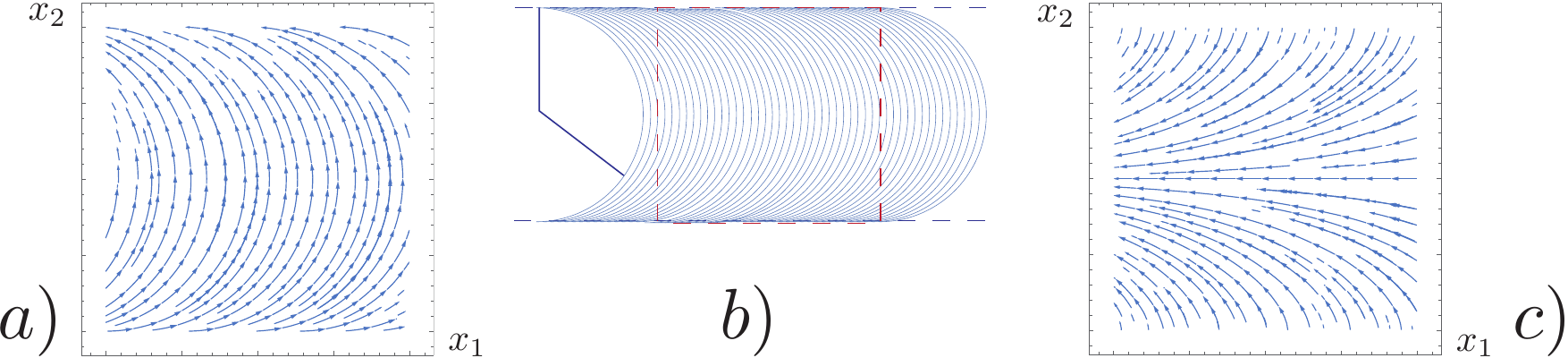}
  \caption{\small \label{fig:patterns} (Color online) $a)$ The director field defined
  by $\psi(x_2)=\cos^{-1}(1-x_2)$ on the square domain $\omega=[0,2]\times[0,2]$.
  This pattern generates constant Gaussian curvature
  $K=\lambda^{-2}-\lambda^{2\nu}>0$ upon stimulation. $b)$ The nematic pattern
  obtained by shifting a semicircular arc of radius
  $R=\frac{1}{\sqrt{K}}\left(\lambda^{-2}-\lambda^{2\nu}\right)^{1/2}$ along
  the $x_1$-axis generates constant positive Gaussian curvature $K>0$ upon
  stimulation. $c)$ The director field defined
  by $\psi(x_2)=\pi/2+\cos^{-1}(1-x_2)$ on the square domain
  $\omega=[0,2]\times[0,2]$.
  This pattern generates constant negative Gaussian curvature
  $K=-\left(\lambda^{-2}-\lambda^{2\nu}\right)<0$ upon stimulation.}
\end{figure*}

We now restrict our attention to director fields of the form 
$\boldsymbol{n}=\cos\psi(x_2)\,\hat{\boldsymbol{e}}^1+\sin\psi(x_2)\,\hat{\boldsymbol{e}}^2$,
whose alignment angle field varies only with respect to one of the coordinates.
The Gaussian curvature upon stimulation is
$
K=-\frac{1}{2}\left(\lambda^{-2}-\lambda^{2\nu}\right)\left(\psi''\sin
2\psi+2\psi'^2\cos 2\psi\right).
$
We can rewrite this as
\begin{equation} 
\frac{d^2}{dx_2^2}\cos2\psi=4\,C(K),
\end{equation} 
where
$C(K)={K}/(\lambda^{-2}-\lambda^{2\nu})$,
and solve for constant $K>0$ to find
\begin{equation}  \label{pattern} 
\psi(x_2)=\pm\frac{1}{2}\cos^{-1}\Big(c_1+c_2\,x_2+2\,C(K)\,x_2^2\Big),
\end{equation}
where $c_1$, $c_2$ are constants of integration. This pattern
generates constant Gaussian curvature $K$ 
wherever it is well-defined.
Now consider the particular solution
$ 
\psi(x_2)=\pm\frac{1}{2}\cos^{-1}\Big(2(1-x_2)^2-1\Big),
$
corresponding to
$c_1=1$, $c_2=-4$ and $K=\lambda^{-2}-\lambda^{2\nu}>0$. This can be rewritten
as
$ 
\psi(x_2)=\cos^{-1}(1-x_2),
$
which describes a well-defined pattern for $0\leq x_2 \leq 2$. The pattern on
the square domain $\omega=[0,2]\times[0,2]$ is shown in
Fig.~\ref{fig:patterns} $a)$.

By integrating
along the director field lines, we notice that the integral curves of this pattern consist of
semicircles of unit radius that are shifted along the $x_1$-axis. Indeed, this
is a specific case of a more general result which we will now discuss. Consider
the director field pattern that is generated by translating the semicircle
$\gamma(t)=(R\cos t,R\sin t)$ (for $-\pi/2 < t < \pi/2$) of radius $R$ along the
$x_1$-direction. It is natural to change coordinates from the Cartesian $(x_1,x_2)$ to
$(t,r)$ where $t$ is the parameter along the curve $\boldsymbol{\gamma}$ and $r$ is 
a new parameter in the direction of translation. That is,
$
x_1(t,r):=\gamma_1(t)+r$ and $x_2(t,r):=\gamma_2(t)$. 
The director field $\boldsymbol{n}=n_1\hat{\boldsymbol{e}}^1+n_2\hat{\boldsymbol{e}}^2$ at each point $(t,r)$ is given by
$
n_1(t)=\frac{\gamma_1'}{\sqrt{\gamma_1'^2+\gamma_2'^2}}$, 
$n_2(t)=\frac{\gamma_2'}{\sqrt{\gamma_1'^2+\gamma_2'^2}}$.

The components 
$
\mathbf{A}=\left((\lambda^2-\lambda^{-2\nu})n_{\alpha}n_{\beta}+\lambda^{-2\nu}\delta_{\alpha\beta}\right)
$
of the metric in Cartesian coordinates transform according to
$\mathbf{A}\rightarrow \mathbf{J}^T\mathbf{AJ}$, where $\mathbf{J}$ is the Jacobian matrix
\begin{equation}
\mathbf{J} = \left(
\begin{array}{cc}
 \partial_t x_1 & \partial_{r}x_1 \\
 \partial_t x_2 & \partial_{r} x_2\\
\end{array}
\right).
\end{equation}
Now a direct computation using the metric compenents with respect to the $(t,r)$
coordinates yields the constant positive Gaussian curvature
\begin{equation}
K=\frac{\lambda^{-2}-\lambda^{2\nu}}{R^2}.
\end{equation}
In particular, if we seek to encode a particular constant positive
Gaussian curvature $K=K_{0}>0$ across an initially flat sheet, we can do so by
encoding the pattern obtained by shifting a semicircle of radius
$R=\frac{1}{\sqrt{K_0}}\left(\lambda^{-2}-\lambda^{2\nu}\right)^{1/2}$ as shown 
in Fig.~\ref{fig:patterns} $b)$.

By the observation that the orthogonal dual of a given 2D director field pattern
generates the exact opposite Gaussian curvature at every point, we can encode 
constant negative Gaussian curvature $K=-K_0$ on a thin nematic sheet by simply
using the orthogonal dual of the pattern that encodes positive curvature
$K_0>0$. Returning to the example of Fig.~\ref{fig:patterns} $a)$, where a pattern
encoding constant positive curvature $K=\lambda^{-2}-\lambda^{2\nu}$ was defined
on the square domain $\omega=[0,2]\times[0,2]$ by
$\psi(x_2)=\cos^{-1}(1-x_2)$, we immediately obtain a pattern on the same domain
which encodes constant negative Gaussian curvature
$K=-\left(\lambda^{-2}-\lambda^{2\nu}\right)$, by simply taking
$\psi(x_2)=\cos^{-1}(1-x_2)+\frac{\pi}{2}$. The resulting pattern is shown in
Fig.~\ref{fig:patterns} $c)$. This pattern is generated by shifting a tractrix curve
along its axis.

For a surface in $\mathbb{R}^3$, the components $a_{\alpha\beta}$ and
$b_{\alpha\beta}$ of the first and second fundamental forms satisfy a system of algebraic
differential equations known as the Gauss-Codazzi-Mainardi equations.
Conversely, any pair $(a,b)$ consisting of a
symmetric and positive definite matrix field $(a_{\alpha\beta})$ and a symmetric
matrix field $(b_{\alpha\beta})$ that satisfy the Gauss-Codazzi-Mainardi
equations determines a unique surface up to a rigid transformation 
in $\mathbb{R}^3$ \cite{ciarlet2005introduction}.
Thus, to determine the
equilibrium configuration of the mid-surface of an initially flat nematic
sheet upon stimulation, we also need to know the
components $b_{\alpha\beta}$ of the second fundamental form that minimize the
bending energy
subject to the Gauss-Codazzi-Mainardi constraints.

For a fixed 2D metric, the problem of identifying equilibrium configurations
that minimize the bending energy reduces to the problem of minimizing 
the \emph{Willmore functional} 
\begin{equation}
I_W=\int_{\omega}H^2\,dS, 
\end{equation}
where $H$ is the mean curvature of the deformed surface,
among isometric immersions of the given metric
\cite{lewicka2011scaling,efrati2011hyperbolic,willmore2012introduction}.
 
For a metric of constant positive Gaussian curvature, it is easy to show that
the Willmore functional is minimized precisely for spherical solutions. That is,
a flat nematic sheet whose director field encodes
constant positive curvature $K$ is expected to form part of a sphere of radius
$R=1/\sqrt{K}$ upon stimulation, assuming that the sheet is small
enough to exclude the possibility of self-intersection \cite{Mostajeran2015}. 
 
In the case of a metric of constant negative Gaussian curvature,
identifying minimizers of the Willmore functional is considerably less
straightforward. In \cite{gemmer2011shape} it is shown that for a hyperbolic
elastic disc that has already undergone local deformations, surfaces that are geodesic discs
lying on hyperboloids of revolution of constant Gaussian curvature are
minimizers of the Willmore functional among \emph{smooth} immersions of the
metric. These solutions will appear as saddle shapes in experiments and are
expected to be energetically favorable for sufficiently small discs.
However, it has been shown numerically that certain non-smooth wavy surfaces
formed as odd periodic extensions of subsets of so-called Amsler surfaces are energetically
more favorable than the smooth saddle shapes that correspond to
discs lying on hyperboloids of revolution when the radius of the hyperbolic disc
is sufficiently large \cite{gemmer2013shape}. We synthesize liquid crystal polymer films with
spatially programmed directors in order to realize shape-changing surfaces that exhibit these phenomena.

The director profile in nematic liquid crystal polymer networks can be programmed
 through a variety of methods, including mechanical and magnetic fields
 \cite{kupfer1991nematic, schuhladen2014iris}. Using these methods, however, it
 is difficult to spatially
 control the director orientation. Here we use nematic networks 
 whose precursors have been specifically designed to align to treated surfaces.
  Using this approach low molar mass nematic liquid crystal monomers are filled 
  between two plates separated by a well-defined gap. The treated surfaces, on 
  the interior of the plates, direct the self-assembly of the liquid crystals 
  along a specific orientation through the thickness of the material. By using 
  reactive nematic mesogens, this director orientation can be trapped in an 
  elastic solid. Spatially complex director patterns in liquid crystal cells 
  are prepared using point-by-point photoalignment of an azobenzene dye by 
  irradiation with polarized light \cite{mcconney2013topography, ware2015voxelated} . By 
  altering the polarization 
  of the incident light, the in-plane orientation of the director of the liquid 
  crystal can be spatially controlled. The resulting director field is a pixelated 
  approximation of the desired smooth pattern with each pixel measuring 100
  $\mu$m $\times$ 100 $\mu$m.
  
\begin{figure}  
\captionsetup{justification=raggedright,
singlelinecheck=false
}
\centering
\includegraphics[width=0.7\linewidth]{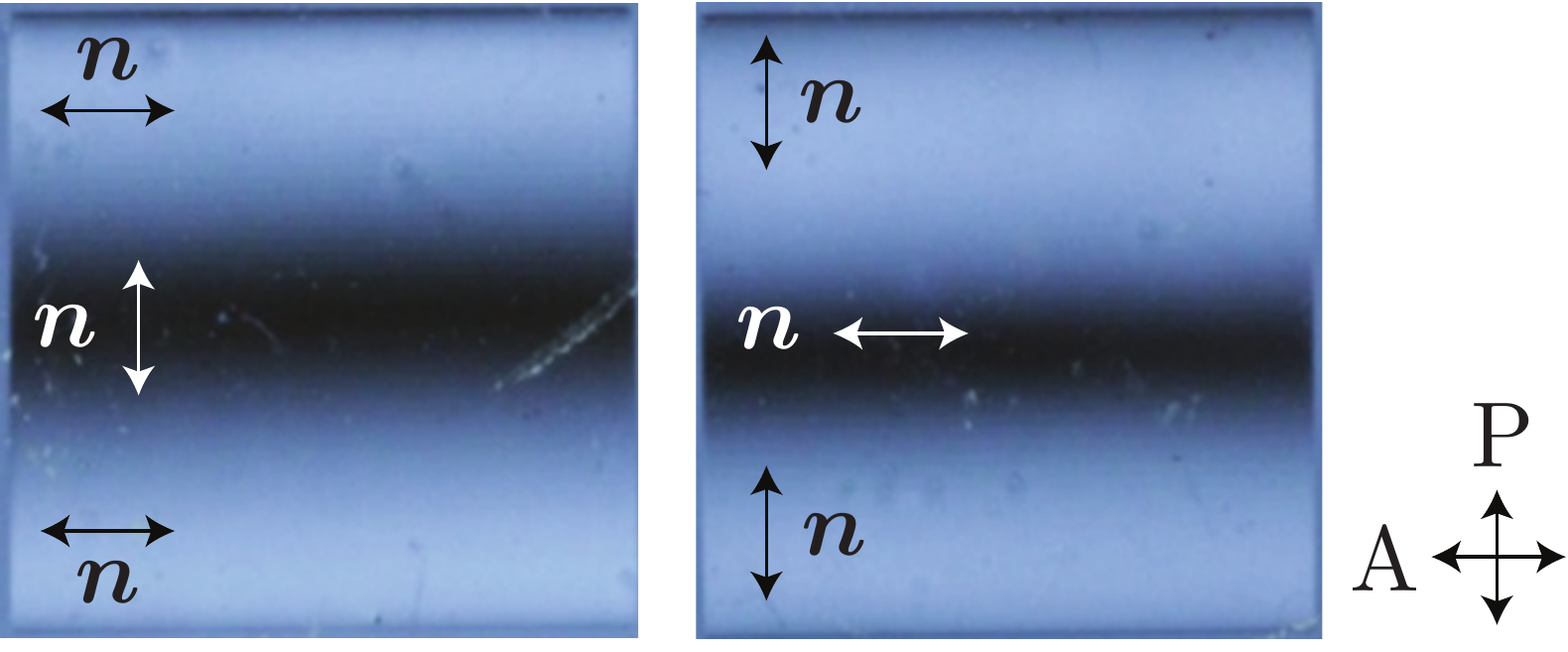}
  \caption{\small (Color online) Polarized optical images of the patterned
  director profiles predicted to generate positive (left) and negative (right) curvature. 
  The patterns are optically equivalent between crossed polarizers. The director 
  orientation at the edges and center of the pattern is indicated with arrows.
  Each square film has a side length of 10 mm.}
  \label{fig:directorProfiles}
\end{figure}

  A number of glassy liquid crystalline polymer networks have been 
  demonstrated to be compatible with surface alignment techniques. Here we use 
  one such composition with $\lambda = 0.94$ and $\nu=0.92$ \cite{wie2015twists}.
  Specifically, we use the composition with the lowest crosslink density from this work. 
  This composition is representative of the larger class of nematic liquid crystal glasses
  that can be aligned using surface alignment techniques \cite{liu2014liquid}.
  The director patterns depicted in Figures 1 $a)$ and 1 $c)$ were chosen to assess the 
  viability of generating Gaussian curvature on exposure to stimulus. After fabrication, the 
  liquid crystal network film is flat at $25^{\circ}C$ and retains the expected birefringence
  of an aligned nematic, as seen in Fig.~\ref{fig:directorProfiles}. 

\begin{figure} 
\captionsetup{justification=raggedright,
singlelinecheck=false
}
\centering
\includegraphics[width=0.9\linewidth]{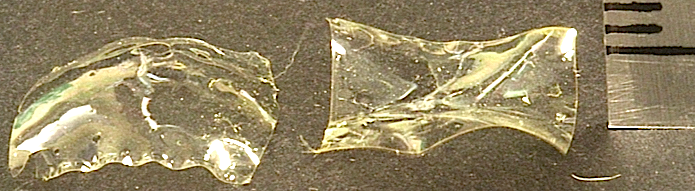}
  \caption{\label{fig:1}\small (Color online) Positive (left) and negative (right) Gaussian
  curvature in $15\,\mathrm{\mu}$m thick glassy liquid crystal network at $175^{\circ}C$.
}
\end{figure}

The thermally-induced shape change
 of 
  the nematic liquid crystal glass is shown in Fig.~\ref{fig:1}.
  As 
  predicted, the pattern depicted in Fig. 1 $a)$ leads to the formation of 
  positive Gaussian curvature, while the pattern from Fig. 1 $c)$ leads to 
  negative Gaussian curvature. On removal of the heat, the film returns to
  a largely flat state. It should be noted that the positive Gaussian 
  curvature sample exhibits a periodic buckling around the edge of the film. 
  This is likely due to the relatively sharp change in director angle  
  with respect to the resolution of the patterning technique near the edges of the film. This buckling
  highlights the limitation on the curvature that can be achieved in nematic liquid crystal
  glasses with comparatively small strains ($\lambda = 0.94$).

\begin{figure*}
\captionsetup{justification=raggedright,
singlelinecheck=false
}
\centering
\includegraphics[width=0.95\linewidth]{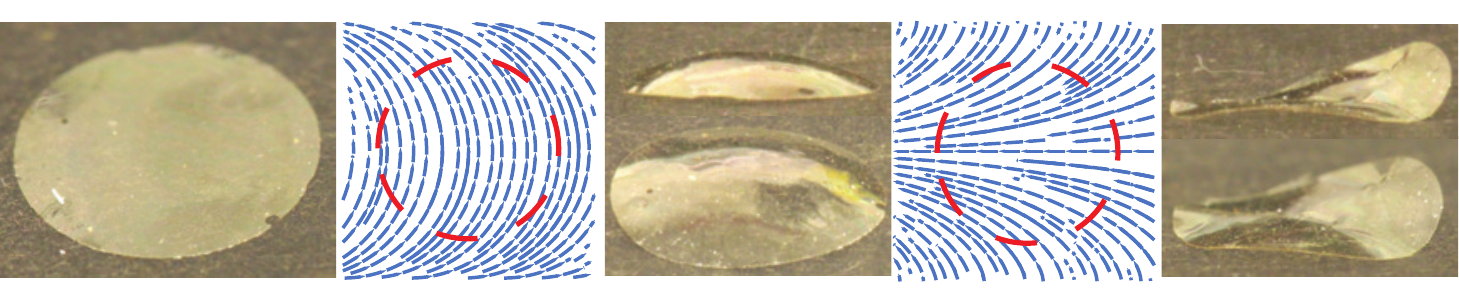}
  \caption{\label{fig:glass4} \small (Color online) From left to right: 1. The
  initially flat configuration of a circular glassy film 15 $\mu$m
  in thickness and 7.1 mm in diameter.
  2.
  The positive Gaussian curvature pattern. The dashed circle indicates the
  boundaries of the circular film. 3. The formation of positive Gaussian
  curvature in the actuated state from two distinct viewing angles.
  4. The negative curvature pattern obtained as the orthogonal dual director
  field. 5. The formation of negative Gaussian curvature in the actuated state from two viewing angles.}
  \label{fig:glass4} 
\end{figure*}

In order to improve the quality of the surfaces that are formed in stimulated nematic glasses, a discoid subsection of the patterned films was removed and exposed to stimulus, as shown in Fig.~\ref{fig:glass4}. In the case of films encoded with either of the identified patterns, the predicted smooth curvature is realized upon stimulation. Indeed,
 shape selection of the equilibrium surface in the positive curvature case seems to be in remarkable
 agreement with the predicted solution of a spherical cap. In the negative curvature case too,
 the equilibrium surface appears largely consistent with the hyperboloid saddle solution that is
 predicted to be energetically favorable for a glassy film at this scale. Due to the tendency for buckling in
 areas of the film where the director changes rapidly with respect to the resolution of the patterning
 process, it should be noted that the curvature
 cannot be increased by simply
 scaling the pattern to smaller dimensions. Instead higher strain materials
 are needed.

\begin{figure}
\captionsetup{justification=raggedright,
singlelinecheck=false
}
\centering
\includegraphics[width=0.75\linewidth]{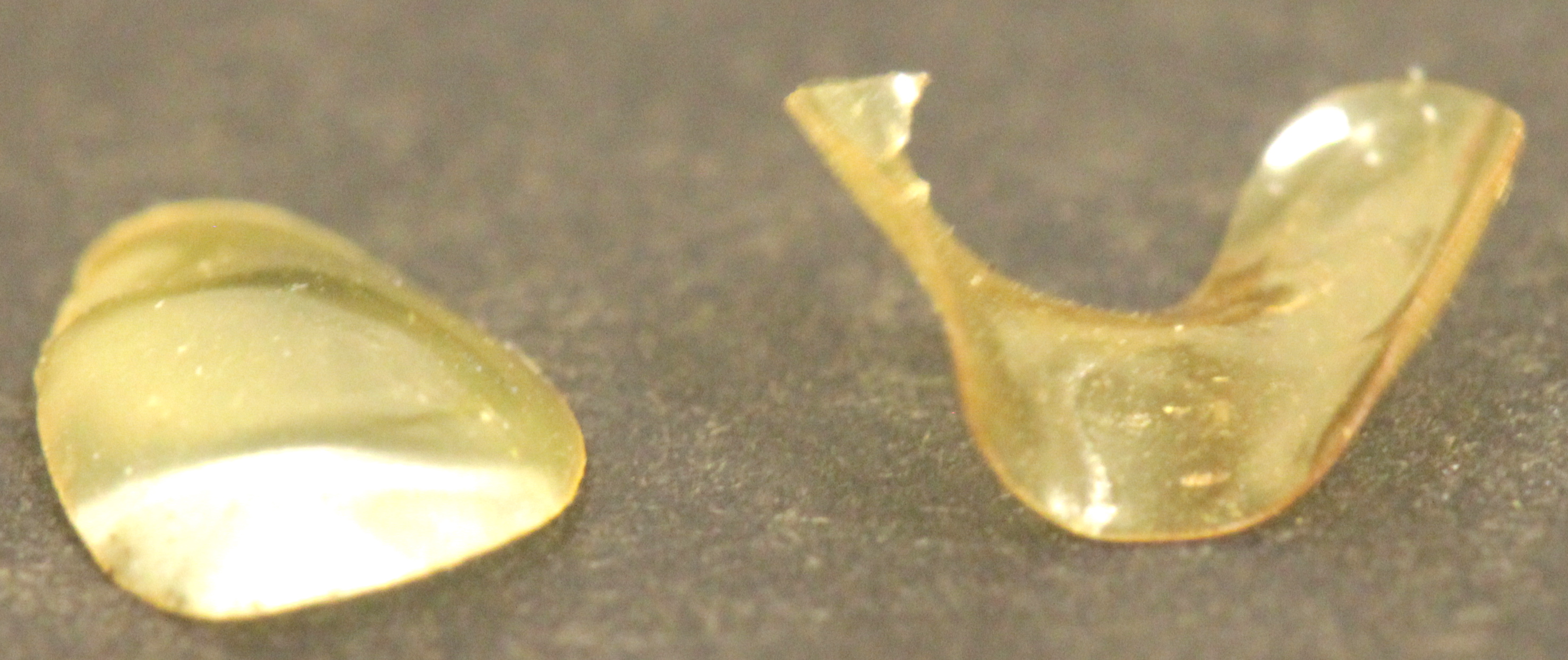}
  \caption{\label{fig:elastomeric}\small (Color online) Side by side comparison
  of positive (left) and negative (right) Gaussian curvature realization in actuated
  elastomeric films.}
\end{figure}

  To facilitate larger curvature realization, we prepare a comparatively high strain 
  surface-alignable liquid crystal elastomer with  $\lambda = 0.65$
  \cite{ware2015voxelated}. 
  Utilizing the pattern depicted in  Fig.~\ref{fig:patterns} $a)$, positive Gaussian curvature is encoded in the
  elastomeric film. As can be seen in Fig.~\ref{fig:elastomeric}, the film encoded for positive Gaussian curvature forms part of a sphere
  with a slightly elliptical distortion. It should be noted that the shape transformation occurs despite the tendency of the 
  director within aligned liquid crystal elastomers to be mobile. This ``soft elasticity" may be contributing to the elliptical 
  distortion of the film. 
  
\begin{figure} 
\captionsetup{justification=raggedright,
singlelinecheck=false
}
\centering
\includegraphics[width=0.75\linewidth]{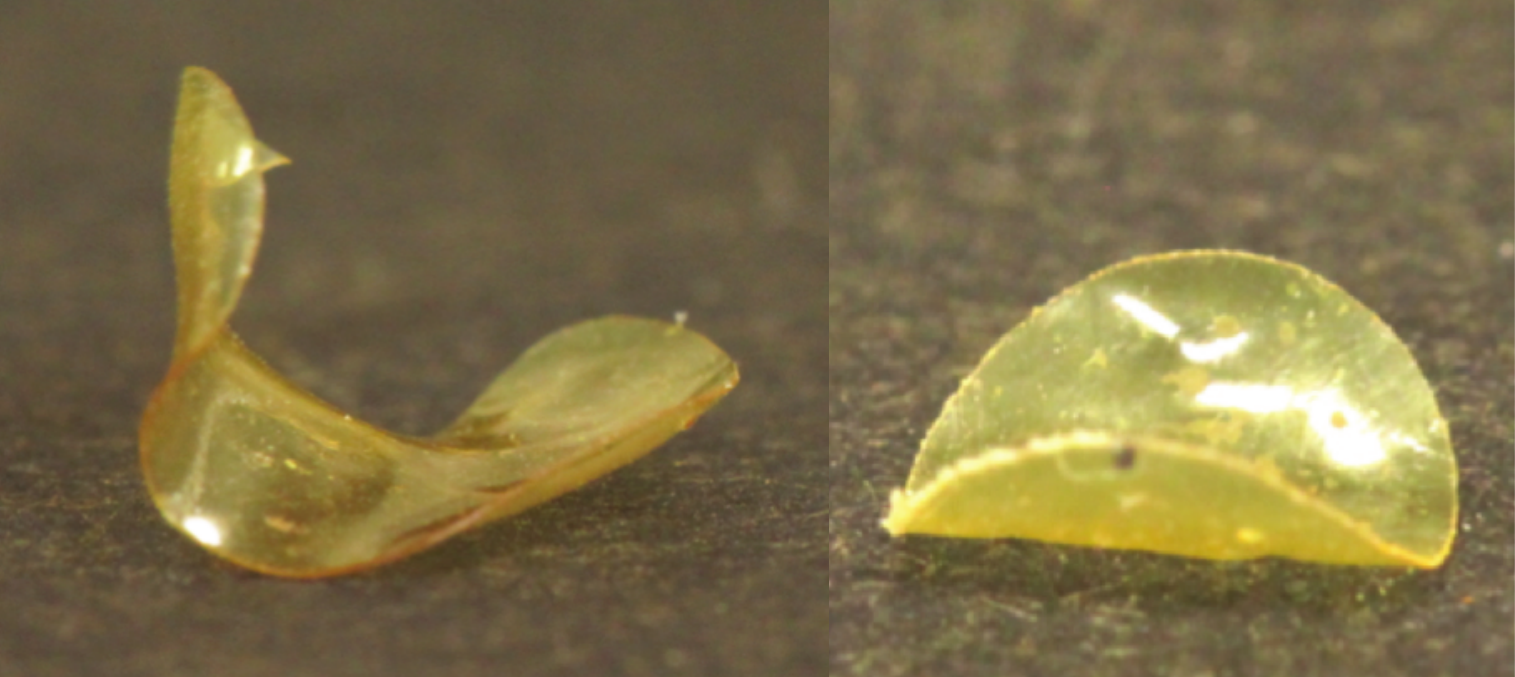}
  \caption{\label{fig:NegativeElastomeric}\small (Color online) Comparison of
  shape selection of discs in the negative Gaussian curvature case depending on
  the size of the domain. For smaller radii (3.7 mm initial diameter) a saddle shape is formed as
  expected (right). When the radius of the disc is sufficiently large (7.1 mm initial diameter), considerably more
  complex surfaces with wavy edges are formed (left).}
\end{figure}

  Fig.~\ref{fig:elastomeric} also shows a complexly buckled hyperbolic surface that is formed when an elastomeric disc encoded with negative Gaussian curvature
  is exposed to stimulus. For smaller diameter films encoded with negative Gaussian curvature, a classic saddle shape can be observed, as shown in 
  Fig.~\ref{fig:NegativeElastomeric}. The surface that is formed by the larger radius hyperbolic disc of Fig.~\ref{fig:NegativeElastomeric} can be interpreted as 
  a distorted periodic Amsler surface. Comparatively, these deformations are significantly
  larger than those observed for the glassy films despite being more than 3 times as thick. Understanding the full spectrum of shape selection for
  films encoded with negative Gaussian
  curvature is an area of ongoing consideration. 

  In summary, our results clearly indicate that Gaussian curvature can be realised in 
  both low-strain, high-modulus glassy and high-strain, low-modulus elastomeric liquid
  crystal networks using appropriate smooth in-plane director fields patterned across initially flat films.  However, our results suggest that
  patterned elastomers are not well suited for potential use in devices which may seek to exploit changing metric geometry to achieve repetitive changes in curvature 
  of thin structures. We conjecture that this is mainly due to the mobility of the directors in such material. On the other hand, our preliminary investigations into patterned nematic   glasses suggest that glassy liquid crystal polymer networks may indeed be promising candidates for use in applications. In particular, the observed shape transformations were in agreement with the theoretical predictions and the behaviour of the films in response to stimulus was found to be robust and reproducible, unlike the case of elastomeric films. We hope that our results will encourage and stimulate further experimental research in achieving desired shape transitions in glassy liquid crystal polymer networks and further work in assessing the viability of their use for specific applications.

\begin{acknowledgments}
CM is supported by the Engineering and Physical
Sciences Research Council of the United Kingdom.
TJW and THW acknowledge financial support from
the Materials and Manufacturing Directorate of the Air Force
Research Laboratory and the Air Force Office of Scientific Research.
The authors are most grateful
to Professor Mark Warner of Cavendish Laboratory for his insightful comments and
support. 

\end{acknowledgments}

\nocite{*}

\bibliography{apssamp}% Produces the bibliography via BibTeX.

\providecommand{\noopsort}[1]{}\providecommand{\singleletter}[1]{#1}%
\begin{thebibliography}{21}%
\makeatletter
\providecommand \@ifxundefined [1]{%
 \@ifx{#1\undefined}
}%
\providecommand \@ifnum [1]{%
 \ifnum #1\expandafter \@firstoftwo
 \else \expandafter \@secondoftwo
 \fi
}%
\providecommand \@ifx [1]{%
 \ifx #1\expandafter \@firstoftwo
 \else \expandafter \@secondoftwo
 \fi
}%
\providecommand \natexlab [1]{#1}%
\providecommand \enquote  [1]{``#1''}%
\providecommand \bibnamefont  [1]{#1}%
\providecommand \bibfnamefont [1]{#1}%
\providecommand \citenamefont [1]{#1}%
\providecommand \href@noop [0]{\@secondoftwo}%
\providecommand \href [0]{\begingroup \@sanitize@url \@href}%
\providecommand \@href[1]{\@@startlink{#1}\@@href}%
\providecommand \@@href[1]{\endgroup#1\@@endlink}%
\providecommand \@sanitize@url [0]{\catcode `\\12\catcode `\$12\catcode
  `\&12\catcode `\#12\catcode `\^12\catcode `\_12\catcode `\%12\relax}%
\providecommand \@@startlink[1]{}%
\providecommand \@@endlink[0]{}%
\providecommand \url  [0]{\begingroup\@sanitize@url \@url }%
\providecommand \@url [1]{\endgroup\@href {#1}{\urlprefix }}%
\providecommand \urlprefix  [0]{URL }%
\providecommand \Eprint [0]{\href }%
\providecommand \doibase [0]{http://dx.doi.org/}%
\providecommand \selectlanguage [0]{\@gobble}%
\providecommand \bibinfo  [0]{\@secondoftwo}%
\providecommand \bibfield  [0]{\@secondoftwo}%
\providecommand \translation [1]{[#1]}%
\providecommand \BibitemOpen [0]{}%
\providecommand \bibitemStop [0]{}%
\providecommand \bibitemNoStop [0]{.\EOS\space}%
\providecommand \EOS [0]{\spacefactor3000\relax}%
\providecommand \BibitemShut  [1]{\csname bibitem#1\endcsname}%
\let\auto@bib@innerbib\@empty
%</preamble>
\bibitem [{\citenamefont {Dervaux}\ and\ \citenamefont
  {Ben~Amar}(2008)}]{dervaux2008morphogenesis}%
  \BibitemOpen
  \bibfield  {author} {\bibinfo {author} {\bibfnamefont {J.}~\bibnamefont
  {Dervaux}}\ and\ \bibinfo {author} {\bibfnamefont {M.}~\bibnamefont
  {Ben~Amar}},\ }\href@noop {} {\bibfield  {journal} {\bibinfo  {journal}
  {Physical Review Letters}\ }\textbf {\bibinfo {volume} {101}},\ \bibinfo
  {pages} {068101} (\bibinfo {year} {2008})}\BibitemShut {NoStop}%
\bibitem [{\citenamefont {Mostajeran}(2015)}]{Mostajeran2015}%
  \BibitemOpen
  \bibfield  {author} {\bibinfo {author} {\bibfnamefont {C.}~\bibnamefont
  {Mostajeran}},\ }\href {\doibase 10.1103/PhysRevE.91.062405} {\bibfield
  {journal} {\bibinfo  {journal} {Phys. Rev. E}\ }\textbf {\bibinfo {volume}
  {91}},\ \bibinfo {pages} {062405} (\bibinfo {year} {2015})}\BibitemShut
  {NoStop}%
\bibitem [{\citenamefont {Klein}\ \emph {et~al.}(2007)\citenamefont {Klein},
  \citenamefont {Efrati},\ and\ \citenamefont {Sharon}}]{klein2007shaping}%
  \BibitemOpen
  \bibfield  {author} {\bibinfo {author} {\bibfnamefont {Y.}~\bibnamefont
  {Klein}}, \bibinfo {author} {\bibfnamefont {E.}~\bibnamefont {Efrati}}, \
  and\ \bibinfo {author} {\bibfnamefont {E.}~\bibnamefont {Sharon}},\
  }\href@noop {} {\bibfield  {journal} {\bibinfo  {journal} {Science}\ }\textbf
  {\bibinfo {volume} {315}},\ \bibinfo {pages} {1116} (\bibinfo {year}
  {2007})}\BibitemShut {NoStop}%
\bibitem [{\citenamefont {Kim}\ \emph {et~al.}(2012)\citenamefont {Kim},
  \citenamefont {Hanna}, \citenamefont {Byun}, \citenamefont {Santangelo},\
  and\ \citenamefont {Hayward}}]{kim2012designing}%
  \BibitemOpen
  \bibfield  {author} {\bibinfo {author} {\bibfnamefont {J.}~\bibnamefont
  {Kim}}, \bibinfo {author} {\bibfnamefont {J.~A.}\ \bibnamefont {Hanna}},
  \bibinfo {author} {\bibfnamefont {M.}~\bibnamefont {Byun}}, \bibinfo {author}
  {\bibfnamefont {C.~D.}\ \bibnamefont {Santangelo}}, \ and\ \bibinfo {author}
  {\bibfnamefont {R.~C.}\ \bibnamefont {Hayward}},\ }\href@noop {} {\bibfield
  {journal} {\bibinfo  {journal} {Science}\ }\textbf {\bibinfo {volume}
  {335}},\ \bibinfo {pages} {1201} (\bibinfo {year} {2012})}\BibitemShut
  {NoStop}%
\bibitem [{\citenamefont {Modes}\ \emph {et~al.}(2011)\citenamefont {Modes},
  \citenamefont {Bhattacharya},\ and\ \citenamefont
  {Warner}}]{modes2011gaussian}%
  \BibitemOpen
  \bibfield  {author} {\bibinfo {author} {\bibfnamefont {C.}~\bibnamefont
  {Modes}}, \bibinfo {author} {\bibfnamefont {K.}~\bibnamefont {Bhattacharya}},
  \ and\ \bibinfo {author} {\bibfnamefont {M.}~\bibnamefont {Warner}},\
  }\href@noop {} {\bibfield  {journal} {\bibinfo  {journal} {Proceedings of the
  Royal Society A: Mathematical, Physical and Engineering Science}\ }\textbf
  {\bibinfo {volume} {467}},\ \bibinfo {pages} {1121} (\bibinfo {year}
  {2011})}\BibitemShut {NoStop}%
\bibitem [{\citenamefont {Aharoni}\ \emph {et~al.}(2014)\citenamefont
  {Aharoni}, \citenamefont {Sharon},\ and\ \citenamefont
  {Kupferman}}]{aharoni2014geometry}%
  \BibitemOpen
  \bibfield  {author} {\bibinfo {author} {\bibfnamefont {H.}~\bibnamefont
  {Aharoni}}, \bibinfo {author} {\bibfnamefont {E.}~\bibnamefont {Sharon}}, \
  and\ \bibinfo {author} {\bibfnamefont {R.}~\bibnamefont {Kupferman}},\
  }\href@noop {} {\bibfield  {journal} {\bibinfo  {journal} {Physical review
  letters}\ }\textbf {\bibinfo {volume} {113}},\ \bibinfo {pages} {257801}
  (\bibinfo {year} {2014})}\BibitemShut {NoStop}%
\bibitem [{\citenamefont {Van~Oosten}\ \emph {et~al.}(2007)\citenamefont
  {Van~Oosten}, \citenamefont {Harris}, \citenamefont {Bastiaansen},\ and\
  \citenamefont {Broer}}]{van2007glassy}%
  \BibitemOpen
  \bibfield  {author} {\bibinfo {author} {\bibfnamefont {C.}~\bibnamefont
  {Van~Oosten}}, \bibinfo {author} {\bibfnamefont {K.}~\bibnamefont {Harris}},
  \bibinfo {author} {\bibfnamefont {C.~W.}\ \bibnamefont {Bastiaansen}}, \ and\
  \bibinfo {author} {\bibfnamefont {D.}~\bibnamefont {Broer}},\ }\href@noop {}
  {\bibfield  {journal} {\bibinfo  {journal} {The European Physical Journal E:
  Soft Matter and Biological Physics}\ }\textbf {\bibinfo {volume} {23}},\
  \bibinfo {pages} {329} (\bibinfo {year} {2007})}\BibitemShut {NoStop}%
\bibitem [{\citenamefont {Warner}\ and\ \citenamefont
  {Terentjev}(2003)}]{warner2003liquid}%
  \BibitemOpen
  \bibfield  {author} {\bibinfo {author} {\bibfnamefont {M.}~\bibnamefont
  {Warner}}\ and\ \bibinfo {author} {\bibfnamefont {E.~M.}\ \bibnamefont
  {Terentjev}},\ }\href@noop {} {\emph {\bibinfo {title} {Liquid crystal
  elastomers}}},\ Vol.\ \bibinfo {volume} {120}\ (\bibinfo  {publisher} {Oxford
  University Press},\ \bibinfo {year} {2003})\BibitemShut {NoStop}%
\bibitem [{\citenamefont {Modes}\ and\ \citenamefont
  {Warner}(2011)}]{modes2011blueprinting}%
  \BibitemOpen
  \bibfield  {author} {\bibinfo {author} {\bibfnamefont {C.~D.}\ \bibnamefont
  {Modes}}\ and\ \bibinfo {author} {\bibfnamefont {M.}~\bibnamefont {Warner}},\
  }\href@noop {} {\bibfield  {journal} {\bibinfo  {journal} {Physical Review
  E}\ }\textbf {\bibinfo {volume} {84}},\ \bibinfo {pages} {021711} (\bibinfo
  {year} {2011})}\BibitemShut {NoStop}%
\bibitem [{\citenamefont {Ciarlet}(2005)}]{ciarlet2005introduction}%
  \BibitemOpen
  \bibfield  {author} {\bibinfo {author} {\bibfnamefont {P.~G.}\ \bibnamefont
  {Ciarlet}},\ }\href@noop {} {\bibfield  {journal} {\bibinfo  {journal}
  {Journal of Elasticity}\ }\textbf {\bibinfo {volume} {78}},\ \bibinfo {pages}
  {1} (\bibinfo {year} {2005})}\BibitemShut {NoStop}%
\bibitem [{\citenamefont {Lewicka}\ and\ \citenamefont
  {Reza~Pakzad}(2011)}]{lewicka2011scaling}%
  \BibitemOpen
  \bibfield  {author} {\bibinfo {author} {\bibfnamefont {M.}~\bibnamefont
  {Lewicka}}\ and\ \bibinfo {author} {\bibfnamefont {M.}~\bibnamefont
  {Reza~Pakzad}},\ }\href@noop {} {\bibfield  {journal} {\bibinfo  {journal}
  {ESAIM: Control, Optimisation and Calculus of Variations}\ }\textbf {\bibinfo
  {volume} {17}},\ \bibinfo {pages} {1158} (\bibinfo {year}
  {2011})}\BibitemShut {NoStop}%
\bibitem [{\citenamefont {Efrati}\ \emph {et~al.}(2011)\citenamefont {Efrati},
  \citenamefont {Sharon},\ and\ \citenamefont
  {Kupferman}}]{efrati2011hyperbolic}%
  \BibitemOpen
  \bibfield  {author} {\bibinfo {author} {\bibfnamefont {E.}~\bibnamefont
  {Efrati}}, \bibinfo {author} {\bibfnamefont {E.}~\bibnamefont {Sharon}}, \
  and\ \bibinfo {author} {\bibfnamefont {R.}~\bibnamefont {Kupferman}},\
  }\href@noop {} {\bibfield  {journal} {\bibinfo  {journal} {Physical Review
  E}\ }\textbf {\bibinfo {volume} {83}},\ \bibinfo {pages} {046602} (\bibinfo
  {year} {2011})}\BibitemShut {NoStop}%
\bibitem [{\citenamefont {Willmore}(2012)}]{willmore2012introduction}%
  \BibitemOpen
  \bibfield  {author} {\bibinfo {author} {\bibfnamefont {T.~J.}\ \bibnamefont
  {Willmore}},\ }\href@noop {} {\emph {\bibinfo {title} {An introduction to
  differential geometry}}}\ (\bibinfo {year} {2012})\BibitemShut {NoStop}%
\bibitem [{\citenamefont {Gemmer}\ and\ \citenamefont
  {Venkataramani}(2011)}]{gemmer2011shape}%
  \BibitemOpen
  \bibfield  {author} {\bibinfo {author} {\bibfnamefont {J.~A.}\ \bibnamefont
  {Gemmer}}\ and\ \bibinfo {author} {\bibfnamefont {S.~C.}\ \bibnamefont
  {Venkataramani}},\ }\href@noop {} {\bibfield  {journal} {\bibinfo  {journal}
  {Physica D: Nonlinear Phenomena}\ }\textbf {\bibinfo {volume} {240}},\
  \bibinfo {pages} {1536} (\bibinfo {year} {2011})}\BibitemShut {NoStop}%
\bibitem [{\citenamefont {Gemmer}\ and\ \citenamefont
  {Venkataramani}(2013)}]{gemmer2013shape}%
  \BibitemOpen
  \bibfield  {author} {\bibinfo {author} {\bibfnamefont {J.}~\bibnamefont
  {Gemmer}}\ and\ \bibinfo {author} {\bibfnamefont {S.~C.}\ \bibnamefont
  {Venkataramani}},\ }\href@noop {} {\bibfield  {journal} {\bibinfo  {journal}
  {Soft Matter}\ }\textbf {\bibinfo {volume} {9}},\ \bibinfo {pages} {8151}
  (\bibinfo {year} {2013})}\BibitemShut {NoStop}%
\bibitem [{\citenamefont {K{\"u}pfer}\ and\ \citenamefont
  {Finkelmann}(1991)}]{kupfer1991nematic}%
  \BibitemOpen
  \bibfield  {author} {\bibinfo {author} {\bibfnamefont {J.}~\bibnamefont
  {K{\"u}pfer}}\ and\ \bibinfo {author} {\bibfnamefont {H.}~\bibnamefont
  {Finkelmann}},\ }\href@noop {} {\bibfield  {journal} {\bibinfo  {journal}
  {Die Makromolekulare Chemie, Rapid Communications}\ }\textbf {\bibinfo
  {volume} {12}},\ \bibinfo {pages} {717} (\bibinfo {year} {1991})}\BibitemShut
  {NoStop}%
\bibitem [{\citenamefont {Schuhladen}\ \emph {et~al.}(2014)\citenamefont
  {Schuhladen}, \citenamefont {Preller}, \citenamefont {Rix}, \citenamefont
  {Petsch}, \citenamefont {Zentel},\ and\ \citenamefont
  {Zappe}}]{schuhladen2014iris}%
  \BibitemOpen
  \bibfield  {author} {\bibinfo {author} {\bibfnamefont {S.}~\bibnamefont
  {Schuhladen}}, \bibinfo {author} {\bibfnamefont {F.}~\bibnamefont {Preller}},
  \bibinfo {author} {\bibfnamefont {R.}~\bibnamefont {Rix}}, \bibinfo {author}
  {\bibfnamefont {S.}~\bibnamefont {Petsch}}, \bibinfo {author} {\bibfnamefont
  {R.}~\bibnamefont {Zentel}}, \ and\ \bibinfo {author} {\bibfnamefont
  {H.}~\bibnamefont {Zappe}},\ }\href@noop {} {\bibfield  {journal} {\bibinfo
  {journal} {Advanced Materials}\ }\textbf {\bibinfo {volume} {26}},\ \bibinfo
  {pages} {7247} (\bibinfo {year} {2014})}\BibitemShut {NoStop}%
\bibitem [{\citenamefont {McConney}\ \emph {et~al.}(2013)\citenamefont
  {McConney}, \citenamefont {Martinez}, \citenamefont {Tondiglia},
  \citenamefont {Lee}, \citenamefont {Langley}, \citenamefont {Smalyukh},\ and\
  \citenamefont {White}}]{mcconney2013topography}%
  \BibitemOpen
  \bibfield  {author} {\bibinfo {author} {\bibfnamefont {M.~E.}\ \bibnamefont
  {McConney}}, \bibinfo {author} {\bibfnamefont {A.}~\bibnamefont {Martinez}},
  \bibinfo {author} {\bibfnamefont {V.~P.}\ \bibnamefont {Tondiglia}}, \bibinfo
  {author} {\bibfnamefont {K.~M.}\ \bibnamefont {Lee}}, \bibinfo {author}
  {\bibfnamefont {D.}~\bibnamefont {Langley}}, \bibinfo {author} {\bibfnamefont
  {I.~I.}\ \bibnamefont {Smalyukh}}, \ and\ \bibinfo {author} {\bibfnamefont
  {T.~J.}\ \bibnamefont {White}},\ }\href@noop {} {\bibfield  {journal}
  {\bibinfo  {journal} {Advanced Materials}\ }\textbf {\bibinfo {volume}
  {25}},\ \bibinfo {pages} {5880} (\bibinfo {year} {2013})}\BibitemShut
  {NoStop}%
\bibitem [{\citenamefont {Ware}\ \emph {et~al.}(2015)\citenamefont {Ware},
  \citenamefont {McConney}, \citenamefont {Wie}, \citenamefont {Tondiglia},\
  and\ \citenamefont {White}}]{ware2015voxelated}%
  \BibitemOpen
  \bibfield  {author} {\bibinfo {author} {\bibfnamefont {T.~H.}\ \bibnamefont
  {Ware}}, \bibinfo {author} {\bibfnamefont {M.~E.}\ \bibnamefont {McConney}},
  \bibinfo {author} {\bibfnamefont {J.~J.}\ \bibnamefont {Wie}}, \bibinfo
  {author} {\bibfnamefont {V.~P.}\ \bibnamefont {Tondiglia}}, \ and\ \bibinfo
  {author} {\bibfnamefont {T.~J.}\ \bibnamefont {White}},\ }\href@noop {}
  {\bibfield  {journal} {\bibinfo  {journal} {Science}\ }\textbf {\bibinfo
  {volume} {347}},\ \bibinfo {pages} {982} (\bibinfo {year}
  {2015})}\BibitemShut {NoStop}%
\bibitem [{\citenamefont {Wie}\ \emph {et~al.}(2015)\citenamefont {Wie},
  \citenamefont {Lee}, \citenamefont {Ware},\ and\ \citenamefont
  {White}}]{wie2015twists}%
  \BibitemOpen
  \bibfield  {author} {\bibinfo {author} {\bibfnamefont {J.~J.}\ \bibnamefont
  {Wie}}, \bibinfo {author} {\bibfnamefont {K.~M.}\ \bibnamefont {Lee}},
  \bibinfo {author} {\bibfnamefont {T.~H.}\ \bibnamefont {Ware}}, \ and\
  \bibinfo {author} {\bibfnamefont {T.~J.}\ \bibnamefont {White}},\ }\href@noop
  {} {\bibfield  {journal} {\bibinfo  {journal} {Macromolecules}\ }\textbf
  {\bibinfo {volume} {48}},\ \bibinfo {pages} {1087} (\bibinfo {year}
  {2015})}\BibitemShut {NoStop}%
\bibitem [{\citenamefont {Liu}\ and\ \citenamefont
  {Broer}(2014)}]{liu2014liquid}%
  \BibitemOpen
  \bibfield  {author} {\bibinfo {author} {\bibfnamefont {D.}~\bibnamefont
  {Liu}}\ and\ \bibinfo {author} {\bibfnamefont {D.~J.}\ \bibnamefont
  {Broer}},\ }\href@noop {} {\bibfield  {journal} {\bibinfo  {journal}
  {Langmuir}\ }\textbf {\bibinfo {volume} {30}},\ \bibinfo {pages} {13499}
  (\bibinfo {year} {2014})}\BibitemShut {NoStop}%
\end{thebibliography}%

\end{document}